\documentclass{article}
\usepackage{ijcai23}

\usepackage{times}
\usepackage{soul}
\usepackage{url}
\usepackage[hidelinks]{hyperref}
\usepackage[utf8]{inputenc}
\usepackage[small]{caption}
\usepackage{graphicx}
\usepackage{amsmath}
\usepackage{amsthm}
\usepackage{booktabs}
\usepackage{algorithm}
\usepackage{algorithmic}
\usepackage[switch]{lineno}

\usepackage{multirow}

\usepackage{svg}

\newtheorem{definition}{Definition}

\usepackage{amssymb}

\newcommand{\M}{\mathcal{M}}

\newcommand{\Z}{\mathbb{Z}}
\newcommand{\src}[1]{{\mathrm{src}(#1)}}
\newcommand{\tgt}[1]{{\mathrm{tgt}(#1)}}

\title{Optimized Crystallographic Graph Generation for Material Science}

\author{
Astrid Klipfel$^{1,2,3}$
\and
Yaël Frégier$^3$\and
Adlane Sayede$^2$\And
Zied Bouraoui$^1$
\affiliations
$^1$Univ. Artois, UMR 8188, Centre de Recherche en Informatique de Lens (CRIL), F-62300 Lens, France.\\
$^2$Univ. Artois, UMR 8181, Unité de Catalyse et de Chimie du Solide (UCCS), F-62300 Lens, France.\\
$^3$Univ. Artois, UR 2462, Laboratoire de Mathématiques de Lens (LML), F-62300 Lens, France.\\
\emails
\{astrid.klipfel,yael.fregier,adlane.sayede,zied.bouraoui\}@univ-artois.fr,
}

\begin{document}

\maketitle

\begin{abstract}
    Graph neural networks are widely used in machine learning applied to chemistry, and in particular for material science discovery. For crystalline materials, however, generating graph-based representation from geometrical information for neural networks is not a trivial task. The periodicity of crystalline needs efficient implementations to be processed in real-time under a massively parallel environment. With the aim of training graph-based generative models of new material discovery, we propose an efficient tool to generate cutoff graphs and k-nearest-neighbours graphs of periodic structures within GPU optimization. We provide pyMatGraph a Pytorch-compatible framework to generate graphs in real-time during the training of neural network architecture. Our tool can update a graph of a structure, making generative models able to update the geometry and process the updated graph during the forward propagation on the GPU side. Our code is publicly available at https://github.com/aklipf/mat-graph.
\end{abstract}

\section{Introdution}


New materials discovery is a fundamental challenge in material sciences where high-throughput screening based on machine learning models is largely employed to obtain materials with desired properties.
Crystalline (crystal) material generation has recently received considerable attention, e.g.\ \cite{xie2021crystal,https://doi.org/10.48550/arxiv.2202.13947,DBLP:conf/aaai/klipfelPHFSB23}. In our setting, we are interested in generating new crystal materials for developing new solar panels with a band gap enabling hydrolyse. This helps to solve problems related to clean energy production and storage, which is one of the major challenges facing our society. It can also be used to produce hydrocarbons from CO2, helping to reduce the carbon footprint of human activities.

From organic chemistry to material science, Graph Neural Networks (GNN) have received increasing attention in a variety of tasks such as classification \cite{SchuttKFCTM17,jorgensen2018neural,gasteiger_dimenet_2020,gasteiger_dimenetpp_2020,doi:10.1021/acs.chemmater.9b01294,Choudhary2021,klicpera2022gemnet} and generation \cite{satorras2021en,xie2021crystal,Long2021,PhysRevMaterials.6.033801,https://doi.org/10.48550/arxiv.2202.13947}. 
Notice that organic molecules are composed of wide carbon chains with a limited variety of atoms, while crystal materials are three-dimensional periodic structures composed of a wide variety of chemical bonds and atoms. The periodic structure of crystals is often represented as a parallelepiped tiling, a.k.a crystal lattice or unit cell. While generating graph-based representations of organic molecules is straightforward, the periodic structure of crystals makes difficult graph processing when training a generative model, and in particular when a massively parallel environment is required. More precisely, generative models may update the geometry of a chemical structure during forward propagation. However, since the graph associated with a given structure is built from the local environment of atoms, a modification of the geometry leads to the modifications of the graph associated with the structure. Consequently, building a generative model with a dynamic graph is hard to achieve on a periodic structure compared to organic molecules. 

When training graph-based generative models for material discovery, cutoff distance is a commonly used technique \cite{SchuttKFCTM17,gasteiger_dimenet_2020,jorgensen2018neural}. It designates a relative distance threshold value above which no interaction between nodes is considered. In the same vein, \cite{jorgensen2018neural,doi:10.1021/acs.chemmater.9b01294} suggests that k-nearest-neighbours (KNN) graphs can also be a good choice for GNN models. KNN-graph is a type of graph where all the nodes are connected to the k-nearest nodes. When processing small molecules, 
any naive strategy of computing the interatomic distances is feasible, allowing to compute KNN or cutoff graph in a short amount of time and reasonable memory. However, for periodic structures which are infinite, the search area should be carefully selected to avoid unnecessary calculation and memory saturation.  In fact, the volume of the search space expands with the cube of the search radius. As such, possible graphs should be generated in milliseconds to be usable in practice during the training process. Moreover, a periodic structure is represented with a multi-graph where a given node can share multiple edges with another and with itself which brings more complexity to the graph generation process. Finally, for big structures, a processing strategy suitable for massively parallel environments should be used in order to deal with a batch of multiple structures at the same time.

To address the aforementioned issues, we propose an efficient tool that solves KNN and cutoff graph generation for crystalline materials. We provide a compatible implementation with PyTorch that performs on GPUs\footnote{Code available at \url{https://github.com/aklipf/mat-graph}}. We used an approach inspired by the KD-tree search algorithm adapted for periodic structures and propose a data structure adapted to massively parallel environments (GPUs) that effectively keeps track of the KNN of each atom. We empirically show the benefits of using our tool.

\section{Crystallographic Graph Generation}
A crystalline structure can be defined with a cloud of atoms and a repetition pattern that represent periodicity. The repetition pattern is often described as a parallelepiped called a lattice or a cell. The periodic structure is obtained with the tiling of the space by the crystal cell. Consequently, a given atom inside of the cell has multiple positions because of the tiling in space and the local environment of an atom which can overlap with adjacent repetition.

\begin{figure}
  \centering
  \includesvg[width=.7\columnwidth]{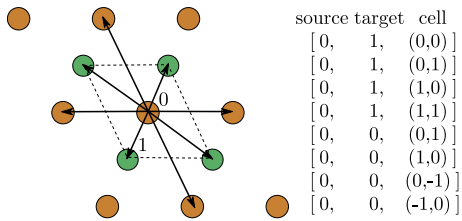}
  \caption{Example of a crystal composed of two atoms, atom 0 is in the centre of the cell and atom 1 is in the bottom left corner. The edges of the graph $\Gamma_1$ associated with a material are composed of the index of a source node, the index of the target node and the coordinate of the cell of the targeted node. }
  \label{fig:graph_material}
\end{figure}

\subsection{Crystallographic Graph}

We follow \cite{DBLP:conf/aaai/klipfelPHFSB23} to define the graph associated with a crystal material as an oriented graph where each edge is represented by triplets containing the index of the source node, the index of the destination node and the relative cell coordinate of the destination node. Figure \ref{fig:graph_material} illustrates this representation. Notice that the definition of graph provided in \cite{DBLP:conf/aaai/klipfelPHFSB23} generalizes to most of the graph definitions proposed in previous works \cite{jorgensen2018neural,doi:10.1021/acs.chemmater.9b01294,satorras2021en}.
%
%
We now recall the formal definition of crystalline structure \cite{DBLP:conf/aaai/klipfelPHFSB23}. 
\begin{definition} 
  The representation space of {\em featured materials} ${\cal M}^F$ is the disjoint union $\coprod_{n \in \mathbb{N}} \M^F_n$ where:
    $$ {\cal M}^F_n =
        \big\{(\rho, x, z) \:|\:  \rho \in GL_d(\mathbb{R}),
        \: x \in [0, 1[^{n \times d},
        \: z \in F^n \big\} $$
    Chemical materials are represented in $\M = \M^{\mathbb{N}}$, with atomic numbers as
    feature sequence $z$.
\end{definition}

$\M^F_n$ is an infinite set of triplet $\rho$, $x$, $z$ that represents all possible materials with $n$ atoms. The atomic number has a chemistry reference, e.g.\ 1 for hydrogen or 6 for carbon. 


\begin{definition}
We call directed 2-graph $\Gamma = (\Gamma_0, \Gamma_1, \Gamma_2)$ 
a triplet of sets together with applications: 
\begin{itemize}
\item $\pi_1 : \Gamma_1 \to \Gamma_0 \times \Gamma_0$, written $\pi_1(\gamma) = (\src{\gamma}, \tgt{\gamma})$
\item $\pi_2 : \Gamma_2 \to \Gamma_0 \times \Gamma_0 \times \Gamma_0$ 
\end{itemize}
We call $\Gamma$ a directed 1-graph when $\Gamma_2 = \varnothing$.
\end{definition}

The aforementioned graphs are often called multi-graphs or hyper-graphs because they generalise 1-graphs
to dimensions $\geq 1$. They are directed because we do not assume any symmetry on $\Gamma$ w.r.t vertice permutations. Recall that $\pi_1$ and $\pi_2$ may not be injective. 

\begin{definition}
\label{def6}
    Let $M = (\rho, x, z)$ in $\M_n^F$ be a material and $c_i > 0$
    for $1 \leq i \leq n$ denotes cutoff distances.
    We define a directed 2-graph
    $\Gamma = \Gamma_{M,c}$ by the graded components: 
    \begin{itemize} 
    \item $\Gamma_0 = \{1, \dots, n \}$ 
    \item $\Gamma_1 = \big\{ (i, j, \tau) \in \Gamma_0 \times \Gamma_0 \times
        \Z^d \: \big|\: || \rho (x_j - x_i + \tau) || < c_i \big\}$ 
    \item $\Gamma_2 = \big\{ (\gamma, \gamma') \in \Gamma_1 \times \Gamma_1
        \: \big|\: \src{\gamma} = \src{\gamma'} \big\}$
    \end{itemize}
\end{definition}

This graph construction includes many definitions of material graphs, making it versatile and usable in most contexts. 
This definition includes a graph built from a constant cutoff distance (i.e. $c_i$ is constant), a graph built from $k$ nearest neighbour or built from chemical properties such as the covalent radii. For more detail, we refer to \cite{DBLP:conf/aaai/klipfelPHFSB23}.

\subsection{Generation Process}
To handle the periodic nature of crystalline, we adapt our graph generation process to work in a torus space. To this end, graph generation is performed by exploring the direct repetition of a cell where we start by evaluating the adjacent cell and extend the search area until we find all the edges. Our graph generation method is built upon two main parts: a searching algorithm and an ordered stack. Combined, the generation process follows an iterative process limiting the RAM usage by splitting the search area. Our generation process remains fast since only a few iterations are required, avoiding useless search areas.

\begin{figure}
  \centering
  \includesvg[width=.7\columnwidth]{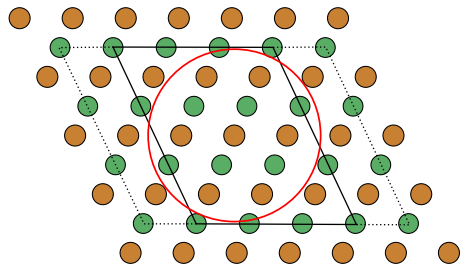}
  \caption{The searching procedure continues while the evaluated area (in the parallelogram with continuous lines) doesn't fully overlap with the search area (represented as a red circle). The next evaluated area (in dotted lines) expends in the direction where the search area is not yet evaluated.}
  \label{fig:searching_algorithm}
\end{figure}

\paragraph{Searching procedure}

Our search procedure is based on a classic KD-tree search strategy.  As shown in Figure \ref{fig:searching_algorithm}, a search radius is used to represent the area where connected nodes can exist. On the other side, we expand the explored area up to a search radius. As the search radius is defined with the KNN in the case of a KNN-graph, the search radius decreases over time when a new area is explored. The search procedure pseudo-code is given by Algorithm \ref{alg:searching_algorithm}.

\begin{algorithm}
    \caption{KNN graph generation algorithm}
    \label{alg:searching_algorithm}
    \textbf{Input}:\\
    $k$: the k nearest connected atoms\\
    $\rho$: the shape of the lattice of the crystal\\
    $x$: the position of the nodes inside the lattice\\
    \textbf{Output}: a set of edges
    \begin{algorithmic}[1] 
        \STATE $d^{\max}_i\gets\infty$
        \STATE $border\gets (0,0,0)$
        \STATE $\Gamma_1\gets \{\emptyset\}$
        \WHILE{$\text{any}(d^{\max}_i>\text{closest\_distance}(border,\rho))$}
        \STATE $extension\gets \text{next\_evaluated\_area}(border,d^{\max}_i)$
        \STATE $\gamma\gets \text{evaluate\_area}(\rho,x,extension)$
        \STATE $\Gamma_1\gets \text{push\_nearest}(\Gamma_1,\gamma,k)$
        \STATE $border\gets border+extension$
        \STATE $d^{\max}_i\gets \max_{(i',j,\tau)\in\Gamma_1|i'=i} d_{i'j\tau}$
        \ENDWHILE
        \STATE \textbf{return} $\Gamma_1$
    \end{algorithmic}
\end{algorithm}

\paragraph{Ordered stack}

To keep track of the k closest points already discovered by our search procedure, we proposed an efficient data structure to store points. Our ordered stack first concatenates new data and then sorts them by distance.
After that, the edges are filtered to keep only the KNN in the case of a KNN-graph or the edges under a given cutoff distance in the case of a cutoff-graph.

\begin{algorithm}
    \caption{Push edges in and ordered stack}
    \label{alg:parallel_ordered}
    \textbf{Input}: \\
    $k$: the k shortest edges\\
    $\Gamma_1$: a list of edges\\
    $\gamma$: the list of edges to merge\\
    \textbf{Output}:  the list of the k shortest edges
    \begin{algorithmic}[1] 
        \STATE $\Gamma'_1\gets\Gamma_1\mathbin\Vert\gamma$
        \STATE $\Gamma'_1\gets\text{sort\_by\_distance}(\Gamma'_1)$
        \STATE $\Gamma'_1\gets\text{stable\_sort\_by\_source\_index}(\Gamma'_1)$
        \STATE $d^k_i\gets\text{k\_nearest\_distance}_i(\Gamma'_1)$
        \STATE $\Gamma'_1\gets\{(i,j,\tau)\in\Gamma'_1|d_{(i,j,\tau)}\leq d^k_i\}$
        \STATE \textbf{return} $\Gamma'_1$
    \end{algorithmic}
\end{algorithm}

In addition to graph generation, our tool provides additional functionalities such as:  
\begin{itemize}
    \item Symmetric graph: as some GNN require symmetric directed graphs to perform specific message-passing schema, our tool includes a procedure that makes a given graph symmetric by adding missing edges while guaranteeing the uniqueness of the edges.

 \item Triplets generation: We provide an implementation to generate triplets composed of two edges sharing the same source nodes during the run-time. This task is important because recent works use triplets information during inference \cite{DBLP:conf/aaai/klipfelPHFSB23,xie2021crystal,klicpera2022gemnet}. 
\end{itemize}

\section{Performance Evaluation}

To evaluate the performance of our tool, we conducted experiments on Materials project \cite{jain2013commentary} which is a dataset composed of 133420 crystalline materials studied with \textit{ab inito} calculation.  We considered the same setting as \cite{xie2021crystal} where structures composed of more than 64 atoms are removed since they are in general considered outliers.  
The experiments are performed on an Nvidia quadro RTX 8000 GPU.

\paragraph{CPU vs GPU}
We compared the time required to process all the structures for our tool with and without GPU optimization. We used a fixed batch size of 256 structures and generated the structures for the 16, the 32 and the 64 nearest neighbours of atoms. As shown in table \ref{tab:total_time}, the KNN-graph generated on GPU is up to 40 times faster than an equivalent CPU library.

\begin{table}
    \centering
    \addtolength{\tabcolsep}{-1.5pt} 
    \resizebox{\linewidth}{!}{\begin{tabular}{rcccccc}
    \toprule   
    &\multicolumn{3}{c}{knn} & \multicolumn{3}{c}{cutoff distance}\\
     & 16 & 32 & 64 & 3.0 & 5.0 & 8.0 \\
     \cmidrule(lr){2-4}
     \cmidrule(lr){5-7}  
      CPU & 1029.6 & 1120.1 & 1669.3 & 968.3 & 1214.4 & 2829.9 \\
        GPU & 25.2 & 25.8 & 37.7 & 20.5 & 34.0 & 57.8 \\
        batch & 0.053 & 0.055 & 0.080 & 0.043 & 0.072 & 0.123\\
        \bottomrule  
    \end{tabular}}
    \caption{Processing time of 119701 filtered structures in seconds for CPU, GPU and Batch configurations. KNN denotes the number of neighbours in the graph while the cutoff distance defines a radius (in Angstrom) inside which all atoms are connected. Batch  corresponds to the generation time for a single batch of 256 structures.}
    \label{tab:total_time}
\end{table}

\paragraph{Complexity, inference time and RAM usage} 
To check the time complexity of our method, we compare the generation time of one batch with various KNN settings in Table \ref{tab:total_time} and batch size in Table \ref{tab:batch_size}). Experiments on batch size have been performed for a KNN-graph with 32 neighbours.

\begin{table}
    \centering
    \resizebox{\linewidth}{!}{\begin{tabular}{rccccc}
    \toprule   
     batch size & 32 & 64 & 128 & 256 & 512 \\
    \midrule
        batch (ms) & 22.0 & 24.7 & 35.0 & 59.3 & 97.5\\
        total (s) & 82.2 & 46.2 & 32.7 & 27.7 & 22.8 \\
        RAM (Mo) & 565.6 & 1079.6 & 2300.1 & 2991.0 & 4811.2\\
    \bottomrule 
    \end{tabular}}
    \caption{{Batch time corresponds to the building time of the graph for a single batch. Total time refers to the time required to process the entire dataset.  RAM to the average GPU RAM used by our graph generation tool.}}
    \label{tab:batch_size}
\end{table}

\section{Conclusion}
We propose an efficient tool to convert crystalline materials into graphs. Our library allows for reducing the time spent during prepossessing. More importantly, the graph conversion is quick enough to be used during the training process without the prepossessing step and updates the graph while updating the geometry of a given structure. Our tool opens new possibilities in generative networks for material science.

\section*{Acknowledgments}
This work has been supported by ANR-22-CE23-0002 ERIANA, ANR-20-THIA-0004 and by HPC resources from GENCI-IDRIS (Grant 2022-[AD011013338]).



\bibliographystyle{named}
\bibliography{ijcai23}

\begin{thebibliography}{}

\bibitem[\protect\citeauthoryear{Chen \bgroup \em et al.\egroup
  }{2019}]{doi:10.1021/acs.chemmater.9b01294}
Chi Chen, Weike Ye, Yunxing Zuo, Chen Zheng, and Shyue~Ping Ong.
\newblock Graph networks as a universal machine learning framework for
  molecules and crystals.
\newblock {\em Chemistry of Materials}, 31(9):3564--3572, 2019.

\bibitem[\protect\citeauthoryear{Choudhary and DeCost}{2021}]{Choudhary2021}
Kamal Choudhary and Brian DeCost.
\newblock Atomistic line graph neural network for improved materials property
  predictions.
\newblock {\em npj Computational Materials}, 7(1):185, Nov 2021.

\bibitem[\protect\citeauthoryear{Ekstr\"om~Kelvinius \bgroup \em et al.\egroup
  }{2022}]{PhysRevMaterials.6.033801}
Filip Ekstr\"om~Kelvinius, Rickard Armiento, and Fredrik Lindsten.
\newblock Graph-based machine learning beyond stable materials and relaxed
  crystal structures.
\newblock {\em Phys. Rev. Materials}, 6:033801, Mar 2022.

\bibitem[\protect\citeauthoryear{Gasteiger \bgroup \em et al.\egroup
  }{2020a}]{gasteiger_dimenetpp_2020}
Johannes Gasteiger, Shankari Giri, Johannes~T. Margraf, and Stephan
  G{\"u}nnemann.
\newblock Fast and uncertainty-aware directional message passing for
  non-equilibrium molecules.
\newblock In {\em Machine Learning for Molecules Workshop, NeurIPS}, 2020.

\bibitem[\protect\citeauthoryear{Gasteiger \bgroup \em et al.\egroup
  }{2020b}]{gasteiger_dimenet_2020}
Johannes Gasteiger, Janek Gro{\ss}, and Stephan G{\"u}nnemann.
\newblock Directional message passing for molecular graphs.
\newblock In {\em International Conference on Learning Representations (ICLR)},
  2020.

\bibitem[\protect\citeauthoryear{Gibson \bgroup \em et al.\egroup
  }{2022}]{https://doi.org/10.48550/arxiv.2202.13947}
Jason~B. Gibson, Ajinkya~C. Hire, and Richard~G. Hennig.
\newblock Data-augmentation for graph neural network learning of the relaxed
  energies of unrelaxed structures.
\newblock arXiv, 2022.

\bibitem[\protect\citeauthoryear{Jain \bgroup \em et al.\egroup
  }{2013}]{jain2013commentary}
Anubhav Jain, Shyue~Ping Ong, Geoffroy Hautier, Wei Chen, William~Davidson
  Richards, Stephen Dacek, Shreyas Cholia, Dan Gunter, David Skinner, Gerbrand
  Ceder, et~al.
\newblock Commentary: The materials project: A materials genome approach to
  accelerating materials innovation.
\newblock {\em APL materials}, 1(1):011002, 2013.

\bibitem[\protect\citeauthoryear{Jørgensen \bgroup \em et al.\egroup
  }{2018}]{jorgensen2018neural}
Peter~Bjørn Jørgensen, Karsten~Wedel Jacobsen, and Mikkel~N. Schmidt.
\newblock Neural message passing with edge updates for predicting properties of
  molecules and materials.
\newblock arXiv, 2018.

\bibitem[\protect\citeauthoryear{Klicpera \bgroup \em et al.\egroup
  }{2021}]{klicpera2022gemnet}
Johannes Klicpera, Florian Becker, and Stephan G{\"u}nnemann.
\newblock Gemnet: Universal directional graph neural networks for molecules.
\newblock In A.~Beygelzimer, Y.~Dauphin, P.~Liang, and J.~Wortman Vaughan,
  editors, {\em Advances in Neural Information Processing Systems}, 2021.

\bibitem[\protect\citeauthoryear{Klipfel \bgroup \em et al.\egroup
  }{2023}]{DBLP:conf/aaai/klipfelPHFSB23}
Astrid Klipfel, Olivier Peltre, Najwa Harrati, Yaël Fregier, Adlane Sayede,
  and Zied Bouraoui.
\newblock Equivariant message passing neural network for crystal material
  discovery.
\newblock In {\em Thirty-Seventh {AAAI} Conference on Artificial Intelligence,
  {AAAI} 2023}. {AAAI} Press, 2023.

\bibitem[\protect\citeauthoryear{Long \bgroup \em et al.\egroup
  }{2021}]{Long2021}
Teng Long, Nuno~M. Fortunato, Ingo Opahle, Yixuan Zhang, Ilias Samathrakis,
  Chen Shen, Oliver Gutfleisch, and Hongbin Zhang.
\newblock Constrained crystals deep convolutional generative adversarial
  network for the inverse design of crystal structures.
\newblock {\em npj Computational Materials}, 7(1):66, May 2021.

\bibitem[\protect\citeauthoryear{Satorras \bgroup \em et al.\egroup
  }{2021}]{satorras2021en}
V\'{\i}ctor~Garcia Satorras, Emiel Hoogeboom, and Max Welling.
\newblock E(n) equivariant graph neural networks.
\newblock In Marina Meila and Tong Zhang, editors, {\em Proceedings of the 38th
  International Conference on Machine Learning}, volume 139 of {\em Proceedings
  of Machine Learning Research}, pages 9323--9332. PMLR, 18--24 Jul 2021.

\bibitem[\protect\citeauthoryear{Schütt \bgroup \em et al.\egroup
  }{2017}]{SchuttKFCTM17}
Kristof Schütt, Pieter-Jan Kindermans, Huziel Enoc~Sauceda Felix, Stefan
  Chmiela, Alexandre Tkatchenko, and Klaus-Robert Müller.
\newblock Schnet: A continuous-filter convolutional neural network for modeling
  quantum interactions.
\newblock In {\em NIPS}, pages 992--1002, 2017.

\bibitem[\protect\citeauthoryear{Xie \bgroup \em et al.\egroup
  }{2022}]{xie2021crystal}
Tian Xie, Xiang Fu, Octavian-Eugen Ganea, Regina Barzilay, and Tommi~S.
  Jaakkola.
\newblock Crystal diffusion variational autoencoder for periodic material
  generation.
\newblock In {\em International Conference on Learning Representations}, 2022.

\end{thebibliography}

\end{document}